\documentclass[a4paper,12pt]{article}
\usepackage[utf8]{inputenc}
\usepackage[T1]{fontenc}
\usepackage{amsmath,amssymb,amsthm,amsfonts}
\usepackage[english]{babel}
\usepackage{enumitem}
\usepackage[all]{xy}
\usepackage{cite}
\usepackage{float}
\usepackage{array}
\usepackage{graphicx}
\usepackage[small]{caption}
\usepackage{subcaption}
\usepackage{authblk}
\usepackage{cmll}
\usepackage{url}
\usepackage[colorlinks = true,linkcolor = blue,urlcolor  = blue,citecolor = blue,anchorcolor = blue]{hyperref}
\usepackage{color,colortbl}
\usepackage{tikz,tikz-cd}
\usepackage[fixlanguage]{babelbib}
\usepgflibrary{shapes.misc}
\usetikzlibrary{shapes.misc}
\usetikzlibrary{patterns,chains,mindmap,arrows,decorations.pathmorphing,decorations.markings,decorations.pathreplacing,backgrounds,positioning,fit,petri,shapes}
\newtheorem{definition}{Definition}
\newtheorem{request}{Query}

\begin{document}
	
	\renewcommand{\labelitemi}{$\bullet$}
	\renewcommand{\baselinestretch}{1}

\title{ICAR, a categorical framework to connect vulnerability, threat and asset managements}
\author[1]{Arnaud Valence}
\affil[1]{\small ESIEA (Graduate school of digital engineering)\\ \texttt{arnaud.valence@esiea.fr}}
\date{}
\maketitle	
\begin{abstract}
	We present ICAR, a mathematical framework derived from category theory for representing cybersecurity NIST and MITRE's ontologies. Designed for cybersecurity, ICAR is a category whose objects are cybersecurity knowledge (weakness, vulnerability, impacted product, attack technique, etc.) and whose morphisms are relations between this knowledge, that make sense for cybersecurity. Within this rigorous and unified framework, we obtain a knowledge graph capable of identifying the attack and weakness structures of an IS, at the interface between description logics, database theory and cybersecurity. We then define ten cybersecurity queries to help understand the risks incurred by IS and organise their defence.
\end{abstract}


\section{Introduction}  

When it comes to cyber systems defense, security operations management has long involved separate tasks: vulnerability management, cyber threat management, and asset management. Today, these disciplines are intended to interoperate within a broader framework, supported by public knowledge bases about cyber threats, vulnerabilities, and IT assets. This interoperability draws an integral and integrated research path, at the interface between ontology language, database theory and cybersecurity, in order to understand how adversaries use vulnerabilities to achieve their goals.

From a general perspective, the research efforts strive to integrate several repositories: the Common Platform Enumeration (CPE) listing IT assets, the Common Vulnerabilities and Exposures (CVE) listing discovered vulnerabilities, the Common Weakness Enumeration (CWE) listing commonly appearing weaknesses, the MITRE ATT\&CK framework listing Adversary Tactics and Techniques (ATT) and the Common Attack Pattern Enumeration and Classification (CAPEC) which helps facilitate attack identification and understanding. The latter repository thus acts as a bridge connecting vulnerability management and threat management. On this basis, research work has explored several avenues.

\begin{itemize}
	\item Some works propose unified ontologies, more or less interoperable, such as Kurniawan et al.\cite{kurniawanATTCKKGLinking2021}, preceded in this by partial ontologies such as UCO and SPESES, which do not yet include the CTI incorporated in the ATT\&CK (even if they can include other vulnerability repositories, such as the CYBOX, KillChain or STUCCO standards).
	\item Other research explores the track of domain-specific languages (DSL), and in essence that of the Meta Attack Language (MAL) meta-language. This is the case for Xiong et al.'s EnterpriseLang meta-language \cite{xiongCyberSecurityThreat2022} and Åberg and Sparf's AttackLang meta-language \cite{abergValidatingMetaAttack2019}.
	\item A third research direction proposes to deepen the graph visualization aspects of attack paths through a relational representation of threats and vulnerabilities. This is the case of the BRON model of Hemberg et al.\cite{hembergLinkingThreatTactics2021}.
\end{itemize}

The approach proposed here is a new way to deepen the mathematical aspects of integrated security operations management. This approach combines three advantages in that

\begin{itemize}
	\item[(i)] like the first approaches mentioned above, it develops a unified vision of vulnerability and threat repositories;
	\item[(ii)] like the second ones, they articulate vulnerabilities and threats within the framework of a cybersecurity-oriented meta-language, except that --- and this is a fundamental point --- it is a \textit{mathematical} meta-language rather than an ontological one\footnote{It may be noted that the DSL approach adds an ontological layer to the ontology already at work in the MITRE and NIST repositories.}.
	\item[(iii)] like the third ones, it deepens the study of graph visualization and structural properties of the unified cybersecurity ontology, by borrowing the powerful and rigorous graph-theoretic concepts of category theory.
\end{itemize}

We believe that category theory can be put to good use by cybersecurity teams. Following the example of a growing number of researchers, involved in more and more diverse fields of knowledge, we believe that the concepts of category theory offer important keys to understanding that simplify and unify the treatment of security operations. We see category theory as the very language of interoperability that enables the integrated management of assets, vulnerabilities, and cyber threats.

The article is organized as follows. The second section discusses the construction of the integrated cybersecurity resource, which will lead to the knowledge graph called ICAR. Based on this, the third section shows how to exploit the knowledge graph to answer different concrete cybersecurity queries. We will see how the categorical concepts allow us to handle bottom-up (from assets to defend to adversaries) as well as top-down (from adversaries to assets) queries. The fourth section concludes.

\section{Building ICAR}   

\subsection{Data sources} 

The data sources are from the knowledge bases provided by the NIST (National Institute of Standards and Technology) and the MITRE Corporation.

\begin{itemize}
	\item Common Platform Enumeration (CPE) is a way of assigning standardized identifiers to classes of IT assets.
	
	\item Common Vulnerabilities and Exposures (CVE) is a knowledge base listing publicly known vulnerabilities. Each CVE entry contains an identification number, a description and at least one reference to publicly known cyber security vulnerabilities. Additional information may include patch information, severity scores and impact assessments according to the Common Vulnerability Scoring System (CVSS), as well as links to exploit information and advisories.
	
	\item Common Weakness Enumeration (CWE) is a knowledge base listing software and hardware weaknesses: flaws, features, breaches, bugs, and other errors in the design, architecture or implementation of software and hardware components that, if left unfixed, can make systems and networks vulnerable to attack. CVE entries have a relational link to CWE entries, as an example of a weakness that actually affects a computer system.
	
	\item Common Attack Pattern Enumeration and Classification (CAPEC) enumerates and classifies attack patterns to facilitate the identification and understanding of attacks. The attack patterns have a tree structure, i.e. they are organised into categories and sub-categories of attacks. They allow the ATTs to be linked to CWE weaknesses.
	
	\item MITRE ATT\&CK framework abstractly describes cyber attack techniques organised into twelve sequential tactics. The framework is presented in a matrix format where the columns represent tactics and the rows represent techniques.
\end{itemize}

These five knowledge bases (or six including CVSS) thus make up an integrated ontological resource for cybersecurity (which we will call ICAR). At this point, it is important to note that this resource only represents the abstract relationships between the data sources. In the language of database, we would say that it shows the column headings of the primary and secondary keys, but not the column entries themselves.

\subsection{Ontologies as knowledge graphs} 

The integrated ontological resource can be represented more formally as a graph.

\begin{definition}[Graph]
	A \emph{graph} $G$ is a sequence $G := (V, E, src, tgt)$, where $V$ et $E$ are sets (respectively the set of vertices and the set of arrows of $G$), and $src,tgt : E \rightarrow V$ are functions (respectively the source and target function of $G$). An arrow $e \in E$ with source $src(e) = v$ and target $tgt(e) = w$ is represented as follows:$$v\xrightarrow{\quad \text{e}\quad}w.$$
\end{definition}

On this basis, it is possible to represent each dictionary (or ontology) by a vertex and each link between dictionaries by an arrow, without forgetting that dictionaries can have internal links. This is the case of CAPEC patterns. For example, the CAPEC-593 pattern (Session Hijacking), linked to the CWE-287 weakness (Improper Authentication) and to several techniques, sub-techniques and MITRE ATT\&CK tactics, has itself children (the CAPEC-60, CAPEC-61, CAPEC-102, CAPEC-107 patterns) and is itself linked to the CAPEC-21 pattern (Exploitation of Trusted Identifiers). We must therefore add to the knowledge graph a loop on CAPEC representing the $\mathsf{ChildOf}$ dependency relation. It is also possible to add the dual relation $\mathsf{ParentOf}$, although redundant, as foreseen by the MITRE corporation. This is also the case for weaknesses. For example, the aforementioned weakness CWE-287 has children CWE-295, CWE-306, CWE-645, CWE-1390, and is itself a child of weakness CWE-284 (Improper Access Control). Finally, it remains to take into consideration the internal structure of the ATT\&CK framework, which is broken down into the dictionaries \textit{Tactics}, \textit{Techniques} (including sub-techniques) and \textit{Procedures}. In this article, we will only deal with tactics and techniques. Sub-techniques will be assimilated to techniques of which they are children. 

\begin{figure}[H]
	\centering
	\begin{tikzcd}
		CVSS & & & \\
		CVE \arrow[r,"Has"] \arrow[u,swap,"Has"] \arrow[d,"Has"] & CWE \arrow[loop,swap,"isChildOf"] \arrow[loop below,distance=5em, in=-135, out=-45,"isParentOf"] \arrow[r,bend left,"Has"] & CAPEC \arrow[loop,swap,"isChildOf"] \arrow[loop below,distance=5em, in=-135, out=-45,"isParentOf"] \arrow[l,bend left,"Has"] \arrow[r,"Has"] & Technique \arrow[loop,swap,"isSubTechniqueOf"] \arrow[d,"accomplishesTactic"]\\
		CPE & & & Tactic
	\end{tikzcd}
	\caption{Representation of the security knowledge graph}
	\label{G}
\end{figure}
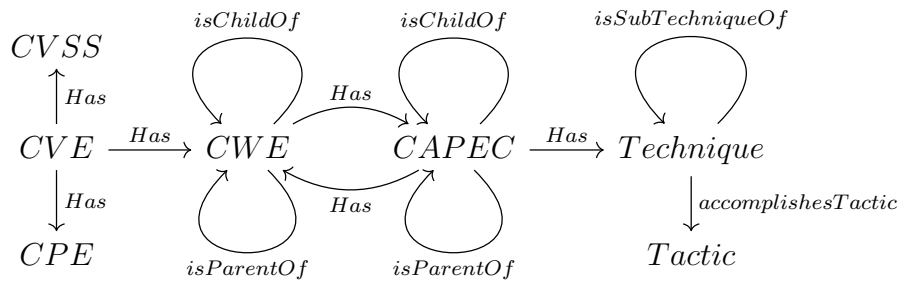

Taking into account these additional specifications, we finally obtain the graph depicted in figure \ref{G}, faithful to the structure of the asset, attack and weakness ontologies. 

Remark the complementarity of the CAPEC and Techniques dictionaries in the overall understanding of threat, beyond their simple logical link. Techniques and (attack) patterns contextualise threat differently. Patterns are intended to focus on the compromise of applications in order to understand the path taken by adversaries to exploit end-to-end application weaknesses in the information system (IS), while techniques describe the concrete dynamics of an attack scenario executed step by step to compromise the IS (see \cite{mitreCAPECATTCK} for more details). Thus, technique T1528, which describes the theft of application access tokens in order to obtain credentials for access to remote systems and resources, can be contextualised in two different ways: (i) as a step to legitimise application actions under the guise of an authenticated user or service by obtaining a trusted identifier, hence its belonging to the CAPEC-21 (Exploitation of Trusted Identifiers) pattern, (ii) as a strategic step to steal account names and passwords, hence its belonging to the TA0006 (Credential Access) tactic. Ultimately, the techniques embody attack tactics as the "how" of the attack (where tactics characterise the "why").

\subsection{Semantic facts and knowledge schema}  

We can do better. What the knowledge graph represents are roughly the data tables (the vertices) and the data columns (the arrows). However, there is still some information missing which is not made explicit in the graph: the path equivalences in $G$.

\begin{definition}[Path]
	Let $G = (V, E, src, tgt)$ be a graph. A \emph{path of length $n$} in $G$, denoted $p \in Path^{(n)}_G$ is a sequence 
	$$p = (v_0 \xrightarrow{\quad a_1 \quad} v_1 \xrightarrow{\quad a_2 \quad} v_2 \xrightarrow{\quad a_3\quad} \dots \xrightarrow{\quad a_n\quad} v_n)$$
	of arrows in $G$. In particular, $Path^{(0)}_G = V$ and $Path^{(1)}_G = E$. The set of all paths in $G$ is denoted $$Path_G := \bigcup_{n\in \mathbb{N}} Path^{(n)}_G.$$
\end{definition}

Paths may themselves carry higher level information about the knowledge structure. This is the case if constraints are imposed on the paths to translate properties that make sense. These constraints can then be expressed as path equivalences.

\begin{definition}[Path equivalence]
	Let $G = (V, E, src, tgt)$ be a graph and $p,q:b\rightarrow c \in Path^{(n)}_G$ two paths in $G$. A \emph{categorical path equivalence relation} in $G$, or simply a path equivalence in $G$, is a relation denoted $\simeq$ such that $p\simeq q$ if and only if $src(p)=src(q)$ and $tgt(p)=tgt(q)$. Moreover, if $m:a\rightarrow b$ and $n:c\rightarrow d$ are two arrows in $G$, then $m$ and $n$ are respectively an epimorphism (a right simplifiable morphism) and a monomorphism (a left simplifiable morphism), i.e. $p\simeq p$ if and only if $mp\simeq mq$ and $pn\simeq qn$.
\end{definition}

Following Spivak\cite{spivakFunctorialDataMigration2012}, we call this equivalence relation \emph{facts}.

There are facts in our study. It is indeed natural to ask for a form of reciprocity in the links between weaknesses and attack patterns. If an attack pattern $CAPEC-X$ exploits a weakness $CWE-Y$, it is natural that it is part of the patterns referenced by this weakness. We can therefore add a path equivalence in the knowledge structure to obtain the following fact:

\begin{equation*} 
	(\mbox{CAPEC-X} \xrightarrow{Has} \mbox{CWE-Y} \xrightarrow{Has}) \simeq \mbox{CAPEC-X}.
\end{equation*}
for all $\mbox{CAPEC-X}\in \mbox{CAPEC}$ and $\mbox{CWE-Y}\in \mbox{CWE}$.

Parent/child relations express other facts. It is natural to require that a weakness \mbox{CWE-X} declaring a child \mbox{CWE-Y} is itself declared as the parent of the child. We therefore have a constraint such as 

\begin{equation*}
	(\mbox{CWE-X} \xrightarrow{isParentOf} \mbox{CWE-Y} \xrightarrow{isChildOf}) \simeq \mbox{CWE-X}.
\end{equation*}

It is also possible to express path equivalences in the more convenient algebraic form of \textit{path equalities}, using composition operator (.). The two previous equivalence relations can then be rewritten, for any $i \in \{\mbox{CWE},\mbox{CAPEC}\}$

\vspace*{-0.5cm}
\begin{equation}
	\begin{split}
		i.\mbox{Has.Has} =& i,\\
		i.\mbox{isChildOf.isParentOf} =& i,\\
		i.\mbox{isParentOf.isChildOf} =& i.\\
	\end{split}
	\label{facts}
\end{equation}

Are there other facts? Could we not ask that the child of a weakness belong to the same attack pattern as its parent, or one of its children? The answer is no. The data structure of the CWE and CAPEC does not have this characteristic. No hybrid facts can be derived from the two previously defined facts.

This negative result can be attributed to the meaning provided by the labelling of arrows. Stress that facts are dependent on the meaning of arrows. \emph{They are semantic facts}. For example, in a bijective data structure where each parent has exactly one child and a child exactly one parent, there is an equivalence between the path $p=(v_0\xrightarrow{isOnlyParentOf} v_1 \xrightarrow{isOnlyChildOf}) \quad \in Path^{(2)}_G$ and the path $p'=v_0 \in Path^{(0)}_G$, but this equivalence no longer holds in a data structure with multiple parents and children.

This is why it is not possible to apply Spivak's theory of ologs\cite{spivakOlogsCategoricalFramework2012}. Ologs are elegant categorical frameworks for rigorously representing knowledge structures exploiting databases, but are limited to structures of functional type. It is inappropriate in this study since the vertices of our knowledge graph generally have several arrows, and may in some cases have none.

On the other hand, the usual theory of oriented (multi)graphs is too broad to capture all the properties present in this study, since we added a path equivalence property. If we add the facts \ref{facts} to the knowledge graph \ref{G}, we obtain a richer structure called (knowledge) \emph{schema} .

\begin{definition}[Categorical schema]
	A categorical schema $S$ consists of a pair $S := (G,\simeq)$ where $G$ is a graph and $simeq$ a path equivalence on $G$.
\end{definition}

In the remainder of this study, we will speak more simply of a schema, in the absence of any risk of confusion.

\subsection{More about relation $CWE \leftrightarrows CAPEC$}  

It was noted that the CWE and CAPEC dictionaries are linked in both directions. This may seem strange, as a mapping can in principle be read both ways: if the weaknesses correctly refer to the attack patterns, it should be possible to recover the former from the latter.

Actually, this is not always the case. Kanakogi et al.\cite{kanakogiTracingCAPECAttack2021} report some CAPEC-IDs that are not identified by CWE-IDs that fall within their attack pattern. As a result, some CVE-IDs would not be correctly mapped to their attack pattern(s). The authors give the example of the CVE-2018-18442 vulnerability, which is linked to a weakness due to network packet flooding. However, while there is an attack pattern for this weakness (the CAPEC-125 pattern), the fact is that the vulnerability is also associated with the CWE-20 weakness (incorrect input validation) which, according to the authors, prevents the vulnerability from being linked to the CAPEC-125 pattern, as the latter is not referenced by the CWE-20 weakness. This problem then motivates the authors to link CVE-IDs directly to CAPEC-IDs. Their solution is to use similarity indicators between CVE-IDs and CAPEC-IDs, using machine learning and natural language processing.

In fact, the traceability problem discussed by Kanakogi et al. does not describe an architectural flaw (since weaknesses can list several attack patterns), but reflects the incomplete mapping between dictionaries. From this point of view, the strategy of the authors seems to be good, even if it consists in directly linking dictionaries that are not graphically related. In the end, this direct approach seems to be complementary to ours in that it allows to complete the collection of arrows that will be used to populate the knowledge schema. This remark is also valid for other approaches of direct mapping between dictionaries, like the projects of Grigorescu et al. \cite{grigorescuCVE2ATTCKBERTBased2022}, Kuppa et al. \cite{kuppaLinkingCVEMITRE2021} or Ampel et al. \cite{ampelLinkingCommonVulnerabilities2021}, which aim to link CVE-IDs to MITRE ATT\&CK \textit{tactics} and \textit{techniques}.

\subsection{ICAR as schema instance}  

The knowledge schema provides an abstract view of cybersecurity data ontologies, the "skeleton". It represents the structure of the data in the form of a triplet (of vertices, arrows and equivalence relations) in exactly the same way as the attributes of database tables present the $n$-uplets of the database. 

It is now a question of populating the knowledge schema in such a way as to make the knowledge base explicit. This explicitation is in fact an \textit{instantiation} (a "concretisation") of its schema.

\begin{definition}[Instance]
	Let $S:=(G,\simeq)$ a categorical schema where $G := (V, E, src, tgt)$ is a graph. An \emph{instance $I$ on} $S$ is given by
	\begin{enumerate}
		\item a set $I(v)$ for any vertex $v\in V$ ;
		\item a function $I(e):I(v)\rightarrow I(v')$ for any arrow $e:v\rightarrow v'$ ;
		\item the equality $I(p)=I(q)$ for any path equivalence $p\simeq q$. 
	\end{enumerate}
\end{definition}

In other words, an instance on $S$ is a path equivalence preserving functor $F:S\rightarrow \mathsf{Set}$.

Among the infinite number of instances that can be generated by $C$, there is one that interests us the most: the up-to-date resource for cybersecurity ontologies. We call this instance \textsf{ICAR} for \textit{Integrated CAtegorical Resource}. To fix ideas, we represent in the tables \ref{sample} an extract of ICAR, where appear at the time of writing the most salient added or updated entries, among more than 20,000 CPE, about 176,000 CVEs, 668 CWEs, 559 CAPECs, 193 Techniques and 14 Tactics.

It is difficult not to make a connection with a database schema, as we suggested above. It is indeed possible to see an arrow $e\in E \in G \in C$ as a relation linking the table identified by $src(e)$ with a table identified by $tgt(e)$. For example, the arrow $\mbox{CWE} \rightarrow \mbox{CAPEC}$ expresses that the table CWE points to the table CAPEC, i.e. entries that have a primary key in CWE are related to entries that have a primary key in CAPEC, via the secondary keys found in the CAPEC column of the table CWE.

At this point we can see that the database schema is not in normal form, since the attribute values are not necessarily atomic (so a weakness frequently has several parents and several CAPECs). Strictly speaking, we should decompose the database schema so as to express it in first normal form. In fact, we do not need such a normalization in this study because it would unnecessarily transform the resource ICAR by adding redundancy. We do, however, need a normal form to check the consistency of ICAR. This leads us to a concept of categorical normal form.

\begin{definition}[Categorical normal form]
	A database is said to be in \emph{categorical normal form} if
	\begin{enumerate}
		\item any table $t$ has a single primary key column $ID_t$ fixed at the beginning;
		\item any entry belonging to a column $c\in t$ refers to a primary key in a single table $t'$, which is denoted by $p_c:t\rightarrow t'$ ;
		\item any database equivalence between two relations $p_c,q_c:t\rightarrow t'$ must be declared as a path equivalence in the corresponding categorical schema, i.e. $p_c\simeq q_c$. 
	\end{enumerate}
\end{definition}

\begin{table}[H] \centering \footnotesize
	\begin{tabular}{|l|}\hline
		\multicolumn{1}{|c|}{CPE} \\\hline
		\textbf{ID}  \\\hline
		\texttt{240.99\_kindle\_books\_project:240.99\_kindle\_books} \\
		\texttt{@nubosoftware/node-static\_project:@nubosoftware/node-static} \\
		\texttt{@thi.ng/egf\_project:@thi.ng/egf} \\
		\texttt{gwa\_autoresponder\_project:gwa\_autoresponder} \\
		\texttt{01org:tpm2.0-tools}  \\\hline
	\end{tabular}
	\quad
	\begin{tabular}{|l||l|l|l|}\hline
		\multicolumn{4}{|c|}{CVE} \\\hline
		\textbf{ID} 	& \textbf{CWE} 	& \textbf{CPE} 			& \textbf{CVSS} \\\hline
		CVE-2023-1684	& CWE-434		& NA* 					& 2.1 \\
		CVE-2023-28371 	& CWE-22		& Stellarium:Stellarium & 4.3 \\
		CVE-2023-21038 	& NA*			& NA* 					& 9.5 \\
		CVE-2023-21039 	& NA*			& NA* 					& 2.1 \\ 
		CVE-2023-21032	& NA*			& NA* 					& 4.1 \\\hline
	\end{tabular}
	\quad
	\begin{tabular}{|l|}\hline
		CVSS\\\hline
		\textbf{ID}  \\\hline
		6.8 \\
		6.9 \\
		7.0 \\
		7.1 \\
		7.2  \\\hline
	\end{tabular}
	\quad
	\begin{tabular}{|l||l|l|l|}\hline
		\multicolumn{4}{|c|}{CWE} \\\hline
		\textbf{ID}	& \textbf{ChildOf} & \textbf{ParentOf}& \textbf{CAPEC} \\\hline
		CWE-787 	& 119 		& 121-124 				& NA* \\
		CWE-79	 	& 74		& 80,81,83-87,692		& 63,85,209,588,591,592 \\
		CWE-89 		& 943		& 564					& 7,66,108-110,470 \\
		CWE-20	 	& 707 		& 179,622,1173,1284-1289& 3,7-10,13,14,22-24... \\
		CWE-125 	& 119		& 126,127				& 540 \\ \hline
	\end{tabular}
	\quad
	\begin{tabular}{|l||l|l|l|l|}\hline
		\multicolumn{5}{|c|}{CAPEC} \\\hline
		\textbf{ID} & \textbf{ChildOf} 	& \textbf{ParentOf} & \textbf{CWE} 	& \textbf{Techniques} \\\hline
		CAPEC-698 	& 542		& -			& 507,829		& 1027,1176,1505,1587 \\
		CAPEC-699 	& 651		& -			& 1300			& 1111 \\
		CAPEC-700 	& 161		& -			& 284			& 1599 \\
		CAPEC-701 	& 94		& - 		& 294,345		& 1557 \\
		CAPEC-702 	& 180 		& -			& 1296			& 1574 \\ \hline
	\end{tabular}
	\quad
	\begin{tabular}{|l||l|}\hline
		\multicolumn{2}{|c|}{Techniques} \\\hline
		\textbf{ID} & \textbf{Tactics} \\\hline
		T1548 		& TA0004,TA0005	 \\
		T1134  		& TA0004,TA0005	 \\
		T1531  		& TA0040		 \\
		T1087  		& TA0007		 \\ 
		T1098 		& TA0003		 \\\hline
	\end{tabular}
	\quad
	\begin{tabular}{|l|}\hline
		Tactics\\\hline
		\textbf{ID}  \\\hline
		TA0043 \\
		TA0042 \\
		TA0001 \\
		TA0002 \\
		TA0003  \\\hline
	\end{tabular}
	\caption{Extracts of ICAR entries. The CPE dictionary is formatted on the following scheme \texttt{cpe:<cpe\_version>:<part>:<vendor>:<product>:<version>: <update>:<edition>:<language>:<sw\_edition>:<target\_sw>:<target\_hw>:<other>}. Only the substring \texttt{<vendor>:<product>} is represented here. ($^\ast$ : Non available)}
	\label{sample}
\end{table}

We check that ICAR actually is in categorical normal form. Condition 1 is met because each dictionary has a single primary key column. Condition 2 is assumed to be met by the successive updates of the dictionaries: if a new entry appears in the foreign key columns, it is assumed that it is indexed at the same time in another table as a primary key. \textit{There are no unreferenced entries in primary key}. On the other hand, it is possible that no foreign key is associated with the entry of a new item as a primary key. This is typically the case when, for instance, an asset affected by a vulnerability has not yet been found, or the weakness corresponding to this vulnerability is still awaiting identification, etc. It is also possible for a primary key column to have no foreign key column. In this case (very common in databases), the table is limited to a single column. This is the case here for the CPE and Tactics tables. In this case, we speak of a leaf column. Condition 3 is respected because it is easy to check that the facts \ref{facts} are translated into relational equivalences in database: the attack patterns declared in the weaknesses declare in turn the declaring weaknesses, and vice versa, and the children declared by the weaknesses or the attack patterns declare in turn their declaring parents.

\section{Using ICAR}   

In this section, we illustrate the applicability of ICAR through several use cases. First of all, we must start by introducing the assets of the IS subject to attack.

\subsection{Instantiate ICAR with an IS} 

Graph \ref{G} brings together knowledge about vulnerability and threat managements in a single categorical schema. But asset management is still to be considered. Assets are explicitly taken into account by Kiesling et al.\cite{kieslingSEPSESKnowledgeGraph2019} in the SEPSES knowledge graph. Indeed, we find there the sub-graph $\mbox{CPE} \xrightarrow{hasProduct} \textsf{Product}$. We take up this idea with two differences. Firstly, we consider only a subset of assets. This restriction allows us to refer to a concrete entity to be analysed, i.e. an IS made up of assets inventoried in a database (to be monitored or investigated). This inventory of assets is commonly materialised by a configuration management database (CMDB). Secondly, and by pure convention, we reverse the arrow formalising the dependency between CPEs and assets. This is indeed what CMDBs suggest, which normally provide for each component added to the database as a primary key a foreign key CPE as illustrated in table \ref{cmdb}.

\begin{table}[H]\centering \small
	\begin{tabular}{|l||l|}\hline
		\multicolumn{2}{|c|}{CMDB} \\\hline
		\textbf{ID} & \textbf{CPE} \\\hline
		A0006 		& \texttt{cpe:2.3:a:microsoft:internet\_explorer:8.0.6001:beta:*:*:*:*:*:*} \\
		VM008  		& \texttt{cpe:2.3:a:vmware:vcenter\_server:6.0:3b:*:*:*:*:*:*} \\
		LB001  		& \texttt{cpe:2.3:h:f5:big-ip\_10250v:-:*:*:*:*:*:*:*}		 \\
		OS007  		& \texttt{cpe:2.3:o:linux:linux\_kernel:2.6.39:*:*:*:*:*:*:*} \\
		OS008 		& \texttt{cpe:2.3:o:paloaltonetworks:pan-os:8.1.16:*:*:*:*:*:*:*} \\	\hline
	\end{tabular}
	\caption{Extract columns ID and CPE from a CMDB}
	\label{cmdb}
\end{table}

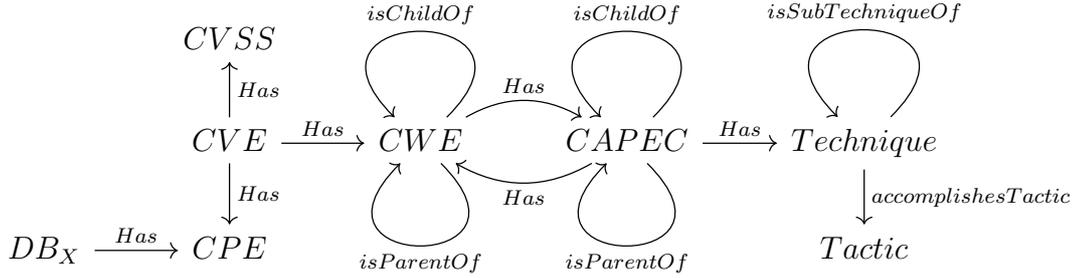
\begin{figure}[H]
	\centering
	\begin{tikzcd}
		& CVSS & & & \\
		& CVE \arrow[r,"Has"] \arrow[u,swap,"Has"] \arrow[d,"Has"] & CWE \arrow[loop,swap,"isChildOf"] \arrow[loop below,distance=5em, in=-135, out=-45,"isParentOf"] \arrow[r,bend left,"Has"] & CAPEC \arrow[loop,swap,"isChildOf"] \arrow[loop below,distance=5em, in=-135, out=-45,"isParentOf"] \arrow[l,bend left,"Has"] \arrow[r,"Has"] & Technique \arrow[loop,swap,"isSubTechniqueOf"] \arrow[d,"accomplishesTactic"]\\
		DB_X \arrow[r,"Has"] & CPE & & & Tactic
	\end{tikzcd}
	\caption{Knowledge schema with inventory of assets}
	\label{G2}
\end{figure}

CMDB can thus be connected to ICAR via the CPE attribute. It can be noted that this correspondance is surjective (each CPE reference refers to at least one asset in the CMDB) but not necessarily injective since a CMDB can have several assets with the same CPE\footnote{It can also happen that the CPE reference is not entered in the CMDB. Furthermore, there are many "exotic" assets that are not listed in the CPE dictionary}. And finally, it is possible to complete the knowledge schema $C$ of which ICAR is the instance, which is represented in figure \ref{G2} by noting $DB_X$ the inventory of assets from the CMDB of SI $X$.

We therefore have the following \textbf{Q1} query:

\begin{request}[\textbf{Q\therequest}]
	Instantiate an inventory of assets $DB_X \in \textsf{Product}$.
\end{request}

We start by noting that the instantiation already referred to here is different from the instantiation of the knowledge schema. The idea now is to instantiate an \textit{object which already has the database structure} (ICAR), in other words to \textit{populate} ICAR (where ICAR instantiates the knowledge schema as a "concretisation"). In category theory, this notion of instantiation can be approached in many ways. In fact, there are at least two ways of dealing with \textbf{Q1}, either by first "connecting" table \textsf{Product} to table CPE and then filtering on the assets $DB_X \subset \textsf{Product}$, or by directly connecting $DB_X$ and CPE tables. In the first case, a filtering operation must be added to the asset connection operation. This operation is not trivial in category theory. Moreover, it implies adding ex post the quantitative aspect induced by the potential presence of more assets with same CPE reference. This is why we will apply the second method, which is easier and more direct. The idea of filtering will nevertheless be discussed later in order to answer query \textbf{Q6}.

In pratical terms, if we think in terms of database management, the addition of $DB_X$ to ICAR can be understood as a database migration, and more precisely as a database union. This intuition can be translated into terms of "categorical data". The idea of "migration" finds a natural translation in category theory with the concept of \textit{functor}. Let $S$ be the (categorical) schema associated with figure \ref{G} (i.e. devoid of assets) and $T$ the schema associated with figure \ref{G2} (i.e. enriched with an inventory of assets). Following the example of Spivak\cite{spivakFunctorialDataMigration2012, spivakRelationalFoundationsFunctorial2015a}, we can then define a schema morphism (i.e. a functor) $F:S \rightarrow T$. Migration functors follow.

\begin{definition}[Migration functors]
	Let $S$ and $T$ be two schemas, $S-\mathsf{Inst}$ and $T-\mathsf{Inst}$ instances on $S$ and $T$ respectively, $F:S \rightarrow T$ a schema morphism and $I\in T-\mathsf{Inst} :T\rightarrow \mathsf{Set}$. Then the composite functor $S \xrightarrow{F} T \xrightarrow{I} \mathsf{Set}$ lives in the $S$-instance ($I\circ F\in S-\mathsf{Inst}$) and we define the functor $\Delta_F$ such that
	\begin{eqnarray}
		\Delta_F : 	& T-\mathsf{Inst} \rightarrow S-\mathsf{Inst}\\
		& I \leadsto I \circ F 
	\end{eqnarray}
	as well as the functors $\Sigma_F, \Pi_F :S-\mathsf{Inst} \rightarrow T-\mathsf{Inst}$ as adjoint functors of $\Delta_F$, respectively on the left and on the right.
\end{definition}

In the language of category theory, $\Delta_F$, $\Sigma_F$ and $\Pi_F$ are called pullback\footnote{Here, the term "pullback" is understood as "a category of instances assigning a set of row-IDs to a schema element". This definition is related to Grothendieck's construction (and fibration).}, left pushforward and right pushforward respectively.

Intuitively, $\Delta_F$ can be understood as a projection operator in the sense that data (tables, columns) is duplicated. In contrast, $\Sigma_F$ is interpreted in terms of unifying tables, and $\Pi_F$ in terms of joining tables. This difference between left and right pushforwards (between unification and junction) is important. When the tables to be joined have no common key, the merging operation can take place in one of two ways: 
\begin{itemize}
	\item either by adding the rows of the second table to those of the first, which has the effect of creating Skolem variables in the unfilled "foreign" columns (in this case we reason on the sum of the primary key spaces);
	\item either by multiplying the rows of the second table with those of the first, which has the effect of duplicating the rows of the first table as many times as there are rows in the second (in this case we reason on the product of the primary key spaces).
\end{itemize}

But the situation is simplified if tables have a common key. In this case, left pushforward and right pushforward are equivalent and there is no duplication of rows or new variables created. This is exactly what happens in our case since asset inventories are supposed to include CPE IDs. Table reconciliation therefore occurs naturally by matching the foreign keys of the inventories with the primary keys of the CPE dictionary.

\subsection{List all vulnerable assets}  

For a CISO or security analyst, one of the most natural queries is to list the vulnerable assets of the IS.

\begin{request}[\textbf{Q\therequest}]
	List all vulnerable assets of a given IS
\end{request}

To process this query, one must first list the entries in the CMDB whose foreign key (i.e. the CPE attribute) also appears as a foreign key in the CVE table. In category theory terminology, we say that we use a pullback (or fiber product), which is one of the many variations of the categorical concept of limit.

\begin{definition}[Pullback]
	Let be the dictionaries CVE and CPE, $DB_X$ the inventory of the IS $X$, and the relations $DB_X \xrightarrow{has} \mbox{CPE}$ and $\mbox{CVE} \xrightarrow{has} \mbox{CPE}$. The pullback of the cospan $DB_X \xrightarrow{has} \mbox{CVE} \xleftarrow{has} \mbox{CPE}$, denoted $DB_X \underset{CPE}\times \mbox{CVE}$, is defined by the set $$DB_X \underset{CPE}\times \mbox{CVE}:=\{(x,y)|x\in DB_X,y\in \mbox{CVE},has(x) = has(y) \}$$
\end{definition}

\noindent respecting the commutative diagram

\begin{figure}[H]
	$$\xymatrixcolsep{3pc}\xymatrix{
		DB_X \underset{CPE}\times \mbox{CVE} \ar[r]^-{has} \ar[d]_{has} & \mbox{CVE} \ar[d]^{has} \\ 
		DB_X \ar[r]_{has} & \mbox{CPE}}$$	
	\caption{Pullback od $DB_X$ and $\mbox{CPE}$} \label{pullback}
\end{figure}

To obtain the only vulnerable assets (dissociated from their vulnerabilities), it is sufficient to retain only the left projection of the pullback. For assets affected by several vulnerabilities, an additional projection morphism is necessary. We then obtain the set of vulnerable assets denoted by $\mathsf{AffectedAssets_X}$.

\subsection{List all vulnerabilities of the IS}  

In the same way, it is also useful to list all vulnerabilities affecting a given IS.

\begin{request}[\textbf{Q\therequest}]
	List all vulnerabilities of a given IS
\end{request}

This query, which is a dual of the previous one, consists in keeping only the vulnerabilities from the pullback \ref{pullback}. This list is obtained by using the right projection of $DB_X \underset{CPE}\times \mbox{CVE}$. The resulting set is denoted $\mathsf{Vuln_X}$.

\subsection{List the vulnerabilities affecting an asset}  

Similarly, it is natural to ask for a list of vulnerabilities affecting a particular asset in the IS.

\begin{request}[\textbf{Q\therequest}]
	List the vulnerabilities affecting an asset $x\in DB_X$.
\end{request}

To process this query, we have to isolate the pairs (asset, vulnerability) of the same asset $x$ in the pullback \ref{pullback}. We therefore need to reason about the following commutative diagram:

$$\xymatrixcolsep{3pc}\xymatrix{
	x \underset{CPE}\times \mbox{CVE} \ar[r]^-{isIn} \ar[d]^{has} & DB_X \underset{CPE}\times \mbox{CVE} \ar[d]^{has} \\ 
	x \ar[r]_{isIn} & DB_X}$$

It turns out that this diagram also defines a pullback, by virtue of the pullback propagation theorem. Consider the following diagram:

$$\xymatrixcolsep{3pc}\xymatrix{
	x \underset{CPE}\times \mbox{CVE} \ar[r]^-{isIn} \ar[d]^{has} & DB_X \underset{CPE}\times \mbox{CVE} \ar[r]^-{has} \ar[d]^{has} & \mbox{CVE} \ar[d]^{has} \\ 
	x \ar[r]_{isIn} & DB_X \ar[r]_{has} & \mbox{CPE}}$$
such that the commutative square on the right-hand side is a pullback. It follows that that of the left-hand side is also a pullback, and consequently the entire commutative diagram. The set $x \underset{CPE}\times \mbox{CVE}$ thus satisfies \textbf{Q4} by providing all vulnerabilities impacting asset $x$.

\subsection{List the assets affected by a vulnerability}  

From the pullback $DB_X \underset{CPE}\times \mbox{CVE}$, we see that it is also possible to filter the resulting pairs by CVE rather than by asset. This filtering fulfils another mission of the CISO (or of any administrator or analyst whether or not they have been mandated to do so): that of monitoring the changes needed to guarantee the logical and physical security of the IS for which he is responsible. This task includes monitoring vulnerabilities likely to affect the IS, and in practice begins by consulting the security alerts issued by the CERT (to which every CISO is in principle a subscriber). Each alert contains one or more CVE entries on a given subject. When a CISO becomes aware of a vulnerability, s/he has to ask her/himself whether the IS is affected, with the level of attention weighted by its CVSS score. Assuming that the new vulnerability is added to ICAR, we therefore have the following \textbf{Q5} query, dual to \textbf{Q4} :

\begin{request}[\textbf{Q\therequest}]
	List the assets affected by a vulnerability $y\in \mbox{CVE}$.
\end{request}

This query is processed by choosing from $DB_X \underset{CPE}\times \mbox{CVE}$ the pairs corresponding to the vulnerability $y$ we are looking for, which we note $DB_X \underset{CPE}\times y$ (i.e. as many pairs as assets impacted by $y$).

The same applies to the resulting commutative diagram, which is a pullback, and by combining \textbf{Q4} with \textbf{Q5} we obtain the pair $(x,y)$ giving the vulnerability $y$ of the asset $x$, that is useful for consulting the remediation status of a vulnerability to be treated (is it fixed, in progress, scheduled...?).

$$\xymatrixcolsep{3pc}\xymatrix{
	x \times y \ar[r]^-{isIn} \ar[d]^{isIn} & DB_X \underset{CPE}\times y \ar[r]^-{has} \ar[d]^{isIn} & y \ar[d]^{isIn} \\ 
	x \underset{CPE}\times \mbox{CVE} \ar[r]^-{isIn} \ar[d]^{has} & DB_X \underset{CPE}\times \mbox{CVE} \ar[r]^-{has} \ar[d]^{has} & \mbox{CVE} \ar[d]^{has} \\ 
	x \ar[r]_{isIn} & DB_X \ar[r]_{has} & \mbox{CPE}}$$

\subsection{List vulnerabilities by criticality}  

In cybersecurity, vulnerabilities are not of equal importance. There is a tendency to focus on the most severe vulnerabilities. It is not uncommon for a CISO to plan enhanced monitoring for critical vulnerabilities. Typically, s/he may request a regular report on vulnerabilities with a score of 9 or more (in CVSS v3.0 notation), or more generally with a score within a range $S \subset [0.0,10.0]$. Query \textbf{Q6} follows.

\begin{request}[\textbf{Q\therequest}]
	List vulnerabilities by CVSS score $s\in S \subset [0.0,10.0]$.
\end{request}

As we saw with \textbf{Q1}, pullback can be used to assign a set of row-IDs to a schema element, which seems to do the trick. However, we need an additional ingredient to filter on the values taken by the entries in the CVSS score column. Indeed, migration functors defined above do not operate in the context of schema morphism, but in that of \textit{type} morphism. We therefore need a notion of \textit{typing}.

\begin{definition}[Typing]	
	Let $S$ be a schema and $A$ a discrete category (i.e. a category containing only objects and identity morphisms) composed of attribute names. A typing for $S$ is a triplet $(A, i, \gamma)$ where $i$ is a functor from $A$ to $S$ mapping each attribute to its vertex, and $\gamma$ is a functor from $A$ to $\mathsf{Set}$, mapping each attribute to its type.
\end{definition}

Then, $i$ reflects the pairing of the knowledge graph's vertices with the attributes of $A$ and $\gamma$ reflects the assignment of the attributes of $A$ to their type. Consequently, we call a \textit{typed instance} a pair $(I, \delta)$ where $I : S \rightarrow Set$ is an instance together with a natural transformation $\delta : I \circ i \Rightarrow \gamma$.

\begin{figure}[H]
	\begin{subfigure}[t]{0.5\textwidth}
		$$\xymatrixcolsep{3pc}\xymatrix{A \ar[rr]^{i} \ar[dr]_{\gamma} & & S \\ & \mathsf{Set} &}$$
		\caption{Typed schema}
	\end{subfigure}
	\begin{subfigure}[t]{0.5\textwidth}
		$$\xymatrixcolsep{3pc}\xymatrix{\ar @{} [drr] |{\Leftarrow_{\delta}} A \ar[rr]^{i} \ar[dr]_{\gamma} & & S \ar[dl]^{I} \\ & \mathsf{Set} &}$$
		\caption{Typed instance}
	\end{subfigure}
	\caption{Typing}
	\label{typing}
\end{figure}

Intuitively, $\delta$ reflects the assignment of a type to each ID in $I$. Typically, this could be the assignment of a \textsf{string} type or a \textsf{float} type, but more generally it can be any type.

Now,  as Spivak points out\cite{spivakFunctorialDataMigration2012}, if we go back to pulback, we see that it is possible to adapt migration functors to \textit{type-change functors}.

\begin{definition}[Type-change functor]
	Let $S$ be a schema and $k : A \rightarrow B$ a morphism of typing instances. We refer to the induced functors $\hat{\Delta}_k : \mathsf{S-Inst}_{/A} \rightarrow \mathsf{S-Inst}_{/B}$ and $\hat{\Sigma}_k, \hat{\Pi}_k : \mathsf{S-Inst}_{/A} \rightarrow \mathsf{S-Inst}_{/B}$ as \textit{type-change functors}. $\hat{\Delta}_k$, $\hat{\Sigma}_k$ and $\hat{\Pi}_k$ are respectively called the pullback, the left pushforward and the right pushforward type-change functor.
\end{definition}

In the context of \textbf{Q8}, we are therefore dealing with a morphism of typing instances which associates a subtype $B$ with the predefined type $A=[0.0,10.0]\supset B$.

\subsection{Measuring the attack surface of an IS} 

The attack surface is a summary of the weak points in a IS that an attacker can exploit to gain access and carry out malicious actions. The more weak points there are, the greater the attack surface and the greater the risk of being attacked. Measuring the attack surface therefore makes it possible to assess the barriers an attacker needs to overcome to exploit the weakness.

\begin{request}[\textbf{Q\therequest}]
	Measuring the attack surface of an IS $X$.
\end{request}

There are myriad ways of defining the attack surface of an IS, and just as many ways of measuring it once it has been defined. One of the simplest definitions is based on the CVSS scores of the vulnerabilities present in the IS. From that point on, the attack surface can be measured in different ways, bearing in mind that the CVSS standard is itself a system of metrics based on three metric groups\cite{NVDVulnerabilityMetrics}\footnote{Base, Temporal, and Environmental. The base metrics produce a score ranging from 0 to 10, which can then be modified by scoring the temporal and environmental metrics. In addition to the base score, the CVSS standard is made up of two other groups of measures: temporal scores and environmental scores. The latter are not provided by the NVD, either because they change over time due to events external to the vulnerability (temporal scores), or because they refer to impacts that are relative to the organisation (environmental scores).}.

The simplest indicators are :
\begin{enumerate}
	\item[(i)] the list of assets affected by a vulnerability with their associated CVSS score (as many weak points exploitable by an attacker) ;
	\item[(ii)] the sum of the assets affected by a vulnerability weighted by their CVSS score.
\end{enumerate}

Formally, indicator (i) corresponds to the set of pairs $\{(DB_X-ID,CVSS-ID)\}$ for any asset $DB_X-ID$ and for any score vulnerability $CVSS-ID$. It is obtained from the schema morphism $CVE \xrightarrow{has} CVSS$ and the pullback $DB_X \underset{CPE}\times CVE$ previously defined as follows:

$$\xymatrixcolsep{3pc}\xymatrix{
	DB_X \underset{CPE}\times CVSS \ar[r]^-{has} & CVSS \\ 
	DB_X \underset{CPE}\times CVE \ar[r]^-{has} \ar[u]_{has} \ar[d]^{has}	& CVE \ar[u]_{has} \ar[d]^{has} \\
	DB_X \ar[r]_{has} & CPE}$$

The product $DB_X \underset{CPE}\times CVSS$ summarises as a simple list the mapping of possible entry points for a potential attacker, with their associated criticality. Seen as the product of $DB_X$ and $CVSS$, $DB_X \times_{CPE} CVSS$ can then be used to define the synthetic indicator (ii). Assuming that the assets affected are of equal importance, the synthetic attack surface indicator, $\mathsf{AttackSurface}$, is easily obtained as the sum of the CVSS scores projected into the list on the right:

$$\mathsf{AttackSurface} := \sum_{(x,y)\in DB_X \underset{CPE}\times CVSS} right(x,y)$$

We note that, despite their equal importance, vulnerable assets do not involve equally important \emph{threats} (such as attack media). Not only do the assets affected differ in the severity of their vulnerabilities, but they can also differ in the number of vulnerabilities affecting them, and it is not uncommon for an asset to accumulate vulnerabilities. For example, Gitlab 15.8.0 has vulnerabilities CVE-2022-3411, CVE-2022-4138, CVE-2022-3759 and CVE-2023-0518, the last three of which are of high severity.

These indicators obviously give a simplistic view of attack surfaces as they actually characterise IT systems. In reality, the assets of an IS do not have the same sensitivity for a variety of reasons: some assets are exposed to the Internet, others are not; some are in production, others in pre-production, development, decommissioning, etc.; some are constrained to high availability, others are not, etc. However, it is possible to take into account the importance of assets by adding a sensitivity criterion. This criterion is generally incorporated into CMDBs, which include a "CI Importance" property for this purpose, in line with ITIL architecture. If affected assets are of unequal importance, then each asset must be weighted by an importance indicator, i.e. a new $IMPT_X$ data set connected to $DB_X$ must be added to ICAR. In this case, it is sufficient to repeat the previous developments by reasoning about the pullback $IMPT_X \times_{CPE} CVSS$ :

$$\xymatrixcolsep{3pc}\xymatrix{
	IMPT_X \underset{CPE}\times CVSS & DB_X \underset{CPE}\times CVSS \ar[l]^-{has} \ar[r]^-{has} & CVSS \\ 
	IMPT_X \underset{CPE}\times CVE \ar[u]_{has} \ar[d]^{has} & DB_X \underset{CPE}\times CVE \ar[l]^-{has} \ar[r]^-{has} \ar[u]_{has} \ar[d]^{has} & CVE \ar[u]_{has} \ar[d]^{has} \\ 
	IMPT_X & DB_X \ar[l]^-{has} \ar[r]_{has} & CPE}$$

Note that an attack surface cannot be interpreted as measures of risk; as the NVD points out\cite{NVDVulnerabilityMetrics}, "CVSS is not a measure of risk". In risk analysis, risk is always the product of a threat, a vulnerability and a severity. ICAR lacks far too much information to be used as a basis for risk analysis, both in terms of business analysis (business values, feared events, impact of damage suffered) and threat analysis (sources of risk, attractiveness of the IT target, etc.). CVSS metrics can only measure the severity of vulnerabilities, which is only one component of risk.

\subsection{List vulnerabilities that can be exploited by a technique or tactic}

We now turn to the long paths to examine how vulnerability management is linked to threat management. This link is bidirectional: top-down and bottom-up. We start with the top-down approach. It is natural to ask what vulnerabilities can be exploited by a given technique pursuing a given tactic. This approach makes it possible to map the dangers corresponding to the different tactical stages of the \textit{kill chain}, which is useful for organisations' defenders, who can prioritise vulnerabilities to be remedied, and for its adversaries, who can investigate their attacks. For example, at the start of an attack, the adversaries apply one or more reconnaissance techniques. They may, for example, target a website or an active directory with the aim of compromising accounts, creating accounts, obtaining capabilities (resource development tactics) or even taking their attack a step further with initial access tactics (remote access to the network, installation of a passive listening system, etc.). The list of vulnerabilities that can be exploited by this tactic can then enable the defender to be more vigilant about the assets that could be targeted by the adversary (i.e. a Wordpress application, an LDAP server, etc.). This knowledge is also useful to the adversaries because it tells them what they should be looking for, an asset or a version number if they already know an asset. So we have request \textbf{Q8}.

\begin{request}[\textbf{Q\therequest}]
	List vulnerabilities that can be exploited by a technique or tactic
\end{request}

There are several ways of dealing with this query. The simplest is probably to observe techniques and tactics as sieves.

\begin{definition}[Sieve]
	Let $v$ be a technique or a tactic. A \emph{sieve} on $v$ is a collection $S$ of morphisms such that : 
	\begin{enumerate}
		\item $e \in S \Rightarrow \mathrm{cod}(e) = v$,
		\item $(e \in S \wedge \mathrm{cod}(f) = \mathrm{dom}(e)) \Rightarrow e \circ f \in S$.
	\end{enumerate}
\end{definition}

In other words, a sieve on an object $A$ in ICAR is a collection of arrows of codomain $v$ closed by precomposition of morphisms in ICAR. However, this definition does not correspond exactly to \textbf{Q8}. On the one hand, the universal aspect of the collection of arrows is missing, as we are looking for the list of \textit{all} vulnerabilities that can be exploited by a technique or tactic. This universality property is provided by the notion of \textit{maximal sieve}.

\begin{definition}[Maximal sieve]
	A sieve $S$ on $v$ is said to be \emph{maximal} (or principal) if it contains all the arrows of codomain $v$. It is denoted $\uparrow v$.
\end{definition}

On the other hand, the resulting sieve has too many arrows, since it includes all the precompositions of $v$-target morphisms. However, what counts for \textbf{Q8} are only the CVE-domain arrows. To subtract the other arrows (i.e. arrows of CWE, CAPEC or Sub-technique domain), we need a notion of \textit{differential sieve}. Let $S$ be a sieve on $v$ in ICAR and $S'$ a sieve on $v$ in ICAR', where ICAR' is the subcategory of ICAR without CVEs.

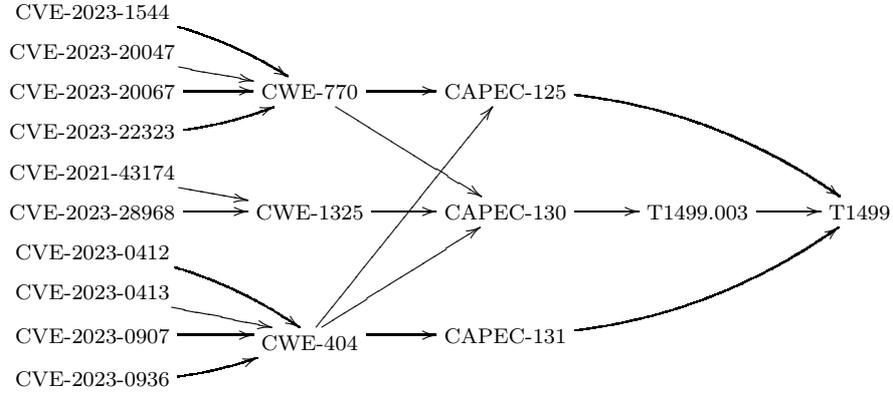
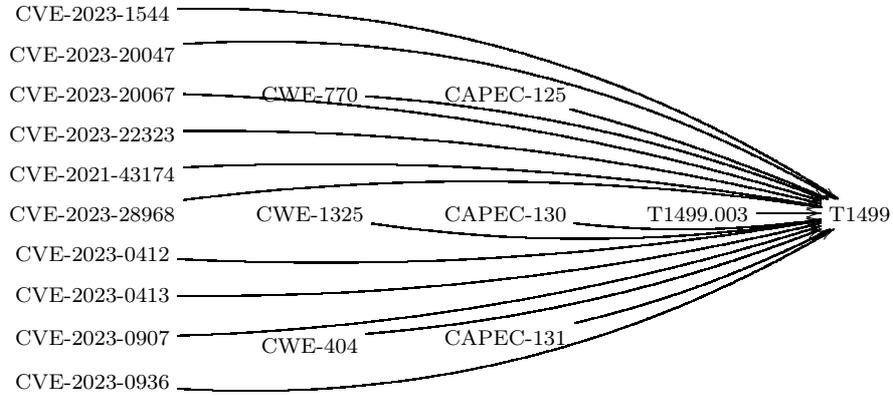
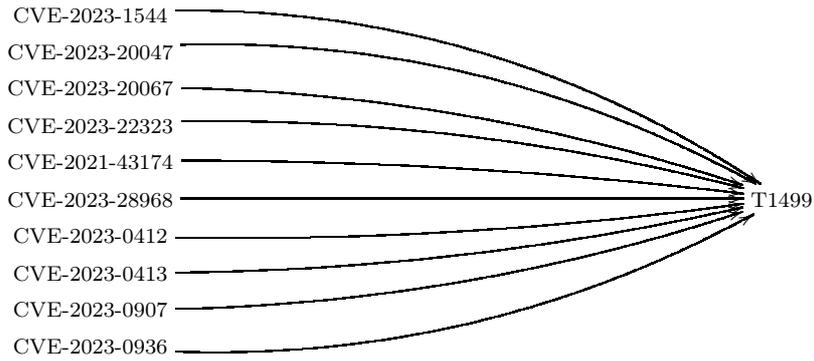
\begin{figure}[H]
	\begin{subfigure}[t]{1\textwidth}
	$$\hspace*{-0cm} \scriptsize \xymatrixcolsep{2pc}\xymatrixrowsep{0.3pc}\xymatrix{
	\mbox{CVE-2023-1544} \ar@/^/[ddr] & & & & \\
	\mbox{CVE-2023-20047} \ar[dr] & & & & \\ 
	\mbox{CVE-2023-20067} \ar[r] &  \mbox{CWE-770} \ar[r] \ar[dddr] & \mbox{CAPEC-125} \ar@/^1pc/[dddrr]  & & \\ 
	\mbox{CVE-2023-22323} \ar@/_/[ur] & & & & \\
	\mbox{CVE-2021-43174} \ar[dr] & & & & \\ 
	\mbox{CVE-2023-28968} \ar[r] & \mbox{CWE-1325} \ar[r] &\mbox{CAPEC-130} \ar[r] & \mbox{T1499.003} \ar[r] & \mbox{T1499}\\ 
	\mbox{CVE-2023-0412} \ar@/^/[ddr]   & & & & \\ 
	\mbox{CVE-2023-0413} \ar[dr]  & & & & \\ 
	\mbox{CVE-2023-0907} \ar[r] & \txt{CWE-404} \ar[uuuuuur]^-{} \ar[uuur] \ar[r] & \mbox{CAPEC-131} \ar@/_1pc/[uuurr] & & \\
	\mbox{CVE-2023-0936} \ar@/_/[ur] & & & & \\ 
	}$$
	\caption{Subcategory extracted under technique T1499.}\label{T1499}
	\end{subfigure}

	\begin{subfigure}[t]{1\textwidth}
	$$\hspace*{-0cm} \scriptsize \xymatrixcolsep{2pc}\xymatrixrowsep{0.3pc}\xymatrix{
	\mbox{CVE-2023-1544} \ar@/^2.2pc/[dddddrrrr] & & & & \\
	\mbox{CVE-2023-20047} \ar@/^2.2pc/[ddddrrrr] & & & & \\ 
	\mbox{CVE-2023-20067} \ar@/^1pc/[dddrrrr] &  \mbox{CWE-770} \ar@/^0.8pc/[dddrrr] & \mbox{CAPEC-125} \ar@/^0.3pc/[dddrr]  & & \\ 
	\mbox{CVE-2023-22323} \ar@/^0.9pc/[ddrrrr] & & & & \\
	\mbox{CVE-2021-43174} \ar@/^0.8pc/[drrrr] & & & & \\ 
	\mbox{CVE-2023-28968} \ar@/^1pc/[rrrr] & \mbox{CWE-1325} \ar@/_0.8pc/[rrr] &\mbox{CAPEC-130} \ar@/_0.5pc/[rr] & \mbox{T1499.003} \ar[r] & \mbox{T1499}\\ 
	\mbox{CVE-2023-0412} \ar@/_0.8pc/[urrrr]   & & & & \\ 
	\mbox{CVE-2023-0413} \ar@/_0.9pc/[uurrrr]  & & & & \\ 
	\mbox{CVE-2023-0907} \ar@/_1pc/[uuurrrr] & \txt{CWE-404} \ar@/_0.8pc/[uuurrr] & \mbox{CAPEC-131} \ar@/_0.5pc/[uuurr] & & \\
	\mbox{CVE-2023-0936} \ar@/_2.2pc/[uuuurrrr] & & & & \\ 
	}$$
	\caption{Maximum sieve $\uparrow T1499$ containing all arrows $e \in ICAR/T1499$ such that $\mathrm{cod}(e)=T1499$ and all arrows $f\circ e$ such that $e \in S$ and $f\in \mathrm{cod}(f)=\mathrm{dom}(e)$.} \label{ST1499} 
	\end{subfigure}

	\begin{subfigure}[t]{1\textwidth}
	$$\hspace*{-0cm} \scriptsize \xymatrixcolsep{4pc}\xymatrixrowsep{0.2pc}\xymatrix{
	\mbox{CVE-2023-1544} \ar@/^2pc/[dddddrrrr] & & & & \\
	\mbox{CVE-2023-20047} \ar@/^1.8pc/[ddddrrrr] & & & & \\ 
	\mbox{CVE-2023-20067} \ar@/^1pc/[dddrrrr] & & & & \\ 
	\mbox{CVE-2023-22323} \ar@/^0.9pc/[ddrrrr] & & & & \\
	\mbox{CVE-2021-43174} \ar@/^0.4pc/[drrrr] & & & & \\ 
	\mbox{CVE-2023-28968} \ar@/^0pc/[rrrr] & & & & \mbox{T1499}\\ 
	\mbox{CVE-2023-0412} \ar@/_0.5pc/[urrrr]   & & & & \\ 
	\mbox{CVE-2023-0413} \ar@/_0.7pc/[uurrrr]  & & & & \\ 
	\mbox{CVE-2023-0907} \ar@/_1pc/[uuurrrr] & & & & \\
	\mbox{CVE-2023-0936} \ar@/_1.8pc/[uuuurrrr] & & & & \\ 
	}$$
	\caption{Differential sieve $S^T$ on technique T1499. The CWE, CAPEC and Sub-technique-domain arrows have disappeared.} \label{DST1499}
	\end{subfigure}
	\caption{Construction stages of \textbf{Q8}}\label{ST}
\end{figure}

In other words, ICAR' consists of the sub-collection of objects from ICAR such that $\mathrm{Ob}(ICAR')=\mathrm{Ob}(ICAR)-\{w\in CVE\}$, and the subcollection of morphisms of ICAR such that $\mathrm{Mor}(ICAR')=\mathrm{Mor}(ICAR)-\{e\in \mathrm{Mor}(ICAR)|src(e)\in CVE\}$. In this context, the object satisfying \textbf{Q8} for techniques is the differential sieve $S^{T}=S\textbackslash S'$. $S^{T}$ therefore contains all the arrows whose domain is the set \textsf{Techniques} and whose codomain is the set \textsf{CVE}. To give a clearer idea, figure \ref{ST} represents the construction stages of \textbf{Q8} for technique T1499 (Endpoint Denial of Service), from the subcategory extracted "under technique T1499" (a), to the maximum sieve on T1499 (b), and finally to the differential sieve (c) answering to \textbf{Q8}.

Obviously, the reasoning is the same for the list of vulnerabilities that can be exploited by a tactic. All we have to do is point the sieve construction $S^{TA}$ to the tactic(s) we want, for example to tactic TA0040 (Impact), which is the tactic performed by technique T1499.

\subsection{List techniques and tactics related to a vulnerability}

We now turn our attention to the bottom-up approach. From the point of view of the defender, it is natural to ask what attack techniques (and therefore tactics) are associated with its vulnerabilities. This knowledge enables him to focus on the vulnerabilities deemed most dangerous from the point of view of their tactical exploitation. This knowledge is also useful for the adversary if he knows some of the targeted assets or even in the absence of any information about the attacked IS. We therefore have query \textbf{Q9}:

\begin{request}[\textbf{Q\therequest}]
	List techniques and tactics related to a vulnerability
\end{request}

This is essentially the dual request of \textbf{Q8}. Since category theory is an ideal framework for studying all kinds of dualities, we just have to do use the dual notions of the two notions defined previously. We thus introduce a notion of \textit{cosieve}.

\begin{definition}[Cosieve]
	Let $v$ be a vulnerability. A \emph{cosieve} on $v$ is a collection $coS$ of morphisms such that : 
	\begin{enumerate}
		\item $e \in coS \Rightarrow \mathrm{dom}(e) = v$,
		\item $(e \in coS \wedge \mathrm{dom}(f) = \mathrm{cod}(e)) \Rightarrow f \circ e \in coS$.
	\end{enumerate}
\end{definition}

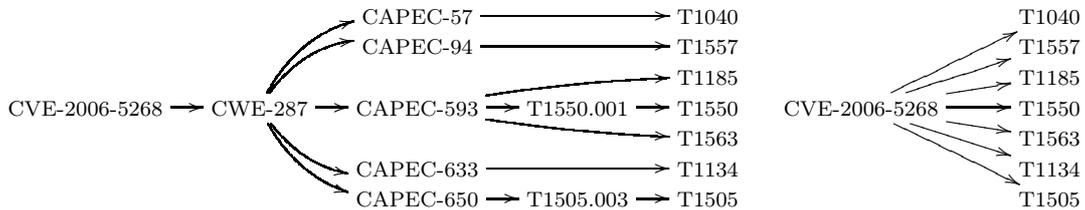
\begin{figure}[H]	
	\centering
	\begin{subfigure}[t]{0.7\textwidth}
		$$\hspace*{-0.4cm} \scriptsize \xymatrixcolsep{1pc}\xymatrixrowsep{0pc}\xymatrix{
		& & \mbox{CAPEC-57} \ar[rr] & & \mbox{T1040} \\ 
		& & \mbox{CAPEC-94} \ar[rr] & & \mbox{T1557} \\
		& & 				    	   & & \mbox{T1185}\\ 
		\mbox{CVE-2006-5268} \ar[r]    & \mbox{CWE-287} \ar@/^1pc/[uuur] \ar@/^1pc/[uur] \ar[r] \ar@/_1pc/[ddr] \ar@/_1pc/[dddr] & \mbox{CAPEC-593} \ar@/^0.2pc/[urr] \ar[r] \ar@/_0.2pc/[drr] &\mbox{T1550.001} \ar[r] & \mbox{T1550} \\ 
		& &					& & \mbox{T1563} \\ 
		& & \mbox{CAPEC-633} \ar[rr] & & \mbox{T1134} \\ 
		& & \mbox{CAPEC-650} \ar[r] & \mbox{T1505.003} \ar[r] & \mbox{T1505} \\ 
		}$$
		\caption{Sub-category of objects and morphisms above CVE-2006-5268.}
	\end{subfigure}\hspace*{0.2cm}
	\begin{subfigure}[t]{0.3\textwidth}
		$$\hspace*{-0cm} \scriptsize \xymatrixcolsep{2pc}\xymatrixrowsep{0pc}\xymatrix{
		& \mbox{T1040} \\ 
		& \mbox{T1557} \\
		& \mbox{T1185} \\ 
		\mbox{CVE-2006-5268} \ar[uuur] \ar[uur] \ar[ur] \ar[r] \ar[dr] \ar[ddr] \ar[dddr] 	& \mbox{T1550} \\ 
		& \mbox{T1563} \\ 
		& \mbox{T1134} \\ 
		& \mbox{T1505} \\ 
		}$$
		\caption{Differential cosieve on CVE-2006-5268}
	\end{subfigure}
	\caption{Construction of \textbf{Q9} for techniques associated with vulnerability CVE-2006-5268}\label{coSTA}
\end{figure}

We then define the notions of maximal cosieve and differential cosieve as before. The differential cosieve $\mathsf{coS}^{TA}$ corresponding to \textbf{Q8} is then given by the complement of the cosieve on $v$ whose target is not a tactic: $\mathsf{coS}^{TA}=\mathsf{C}_{\mathsf{coS}}\mathsf{coS'}= \mathsf{coS}\textbackslash \mathsf{coS'}$, where $\mathsf{coS}$ and $\mathsf{coS'}$ are cosieve on a vulnerability $v$ in ICAR and ICAR' respectively. $\mathsf{coS}^{TA}$ is the collection of arrows with source $v$ and target $Tactics$. The construction is the same for techniques. Simply define the set $ICAR''=ICAR'-Techniques$ and repeat the reasoning from the cosieves in ICAR' and ICAR''. Figure \ref{coSTA} depict the construction of the differential cosieve on vulnerability CVE-2006-5268 (administrative access to the RPC interface) for techniques, from (a) the sub-category of objects and morphisms above CVE-2006-5268 to (b) the final differential cosieve on CVE-2006-5268 satisfying \textbf{Q9}.

\subsection{Measuring the threat surface of an IS} 

The "threat surface" is the set of techniques (or tactics) that an attacker can use to exploit the vulnerabilities of an IS. The threat surface is the counterpart of the attack surface on threat management\footnote{The threat surface is strictly speaking an attack surface, but since this name is usually used to describe the vulnerabilities of the IS, we use the term "threat surface"}.

\begin{request}[\textbf{Q\therequest}]
	Measuring the threat surface of IS $X$
\end{request}

Formally, the threat surface is a simple extension of the differential cosieve used to list the techniques and tactics associated with a vulnerability. We just need to apply the differential cosieve to all the vulnerabilities in the IS, i.e. to the set $\mathsf{Vuln}_X$.

\section{Conclusion and future work}  

The aim of this article was to provide a mathematical foundation for common queries in cybersecurity management. The proposed ICAR categorical model thus covers vulnerability management, threat management and asset management in a unified framework. However, ICAR is not a method for enriching cybersecurity ontologies. In particular, it does not allow the investigation of relations between vulnerability management and threat management. In this sense, the empirical results of the queries examined here are dependent on the quality of the data they use. Our model therefore underlines the importance of work aimed at more finely meshing the various dictionaries of the NIST and the MITRE corporation. Generally speaking, it is clear that query and visualisation models will be enhanced by AI-based works mentioned above.

This article only gives an overview of possible queries for cybersecurity operations. Others could naturally have been envisaged, such as the search for the shortest attack path (i.e. the path with the fewest breaches to exploit). Other queries will be considered later on. Future work will also address the algorithmic design of queries. In this sense, ICAR model should also be seen as a mathematical foundation for establishing a database schema compatible with the defined categorical schema and associated categorical notions. In other words, the queries dealt with in this article will subsequently be extended in terms of query language (SQL), with the aim of providing a bidirectional dictionary between conceptual categorical queries and database queries.

\bibliography{icar} 
\bibliographystyle{plain}

\end{document}